\documentclass[11pt]{article}

\newcommand{\reportversion}{v1.0}

\usepackage[T1]{fontenc}
\usepackage[utf8]{inputenc}

\usepackage[margin=1in]{geometry}
\usepackage{setspace}
\usepackage{titlesec}

\usepackage{amsmath}
\usepackage{amssymb}
\usepackage{mathtools}
\usepackage{bm}
\usepackage{enumitem}
\usepackage{graphicx}
\usepackage{caption}
\usepackage{subcaption}
\usepackage{booktabs}
\usepackage{multirow}
\usepackage{float}
\usepackage{tabularx}
\usepackage{xcolor}
\usepackage{siunitx}
\usepackage{microtype}
\usepackage{algorithm}
\usepackage{algpseudocode}

\usepackage{tikz}
\usetikzlibrary{
  intersections,
  arrows.meta,
  positioning,
  calc,
  shapes.geometric,
  shapes.misc,
  shapes.symbols,
  fit,
  backgrounds,
  decorations.pathreplacing
}

\usepackage{pgfplots}
\usepgfplotslibrary{fillbetween}
\pgfplotsset{compat=1.18}

\usepackage{hyperref}

\titleformat{\section}{\large}{\thesection}{1em}{}
\titleformat{\subsection}{\normalsize}{\thesubsection}{1em}{}

\titlespacing*{\section}{0pt}{1.5ex}{0.8ex}
\titlespacing*{\subsection}{0pt}{1.2ex}{0.6ex}

\setlist[itemize]{leftmargin=*, itemsep=2pt, topsep=4pt}
\setlist[enumerate]{leftmargin=*, itemsep=2pt, topsep=4pt}

\hypersetup{
    colorlinks=true,
    linkcolor=blue,
    citecolor=blue,
    urlcolor=blue,
    pdfauthor={Michael Luby},
    pdftitle={ARC: Consistent, Low-Latency Delivery via Receiver-Side Scheduling}
}

\newcommand{\reportnumber}{BR-TR-2026-02}
\newcommand{\reporttitle}{ARC: Consistent, Low-Latency Delivery \\ via Receiver-Side Scheduling}
\newcommand{\reportauthors}{Michael Luby}
\newcommand{\reportaffiliation}{BitRipple, Inc.}
\newcommand{\reportdate}{May 4, 2026}

\newcommand{\arc}{ARC}
\newcommand{\arcq}{ARC-Q}
\newcommand{\arcqg}{ARC-QG}

\title{\reporttitle}

\author{
    \reportauthors\\
    \reportaffiliation\\
    Berkeley, CA
}

\date{
    \reportdate\\[0.5em]
    {\small \reportnumber\ (\reportversion)}
}

\begin{document}

\maketitle


\begin{abstract}

Applications such as cloud gaming, video streaming, telemetry, ML inference,
and data transfer provide a better experience when data is released at the receiver with timing
reflecting how the data enters the sender. In practice, network delay
variation and recovery dynamics at the receiver distort this timing even when
transports deliver all packets correctly, producing visible jitter, stalls,
and unstable playback. Many such applications operate best when delivery
preserves this timing behavior and its implied order; out-of-order or
irregular delivery can significantly degrade performance even when all data
eventually arrives.

We present a lightweight receiver-side release scheduling protocol, Adaptive
Release Control (\arc), that restores this timing at the receiver. \arc\
releases recovered data in a manner that follows the sender's timing,
maintaining ordering and limiting reordering when necessary while producing
smooth delivery with minimal added latency given network conditions. It
operates entirely on the receiver clock and requires no feedback,
synchronization, or changes to the underlying transport.

As an example, we integrate \arc\ into LT3, a network-layer system currently
deployed as a software overlay that forwards traffic without altering the
transport protocols it carries, where \arc\ functions as an independent
module that regulates release timing for forwarded data.

Evaluating LT3 with \arc\ on a cloud-gaming workload shows that the protocol
removes virtually all large jitter excursions and yields release intervals
that closely match the sender's timing, translating into improved perceptual
smoothness. Broader latency improvements arise from the behavior of the full
LT3 system. The benefits of \arc\ extend to transport protocols carried over
LT3, including TCP, QUIC, WebRTC, UDP, and RTP, as preserving sender timing
improves their behavior across a wide range of conditions.

\end{abstract}

\section{Introduction}

Applications such as cloud gaming, video streaming, telemetry, ML inference, and data transfer provide a better experience when data arrives with timing reflecting the sender-side timing of the data stream. In practice, network delay variation and recovery dynamics at the receiver distort this timing even when transports deliver all packets correctly, producing jitter, stalls, and unstable behavior. These effects arise because recovery at the receiver occurs with irregular timing, which directly propagates to the application when data is released immediately upon arrival.

In systems such as LT3~\cite{AggarwalLubyMinder2025Immersive}, incoming packets are grouped into data blocks for encoding and transmission, and these blocks are reconstructed at the receiver before being released. Variation in packet arrival times causes the recovery time of each reconstructed block to fluctuate, so even when every packet is delivered correctly the receiver observes irregular recovery times.  Small fluctuations in these recovery times translate directly into jitter, buffer oscillation, and head-of-line stalls. Because recovery often occurs in bursts or with uneven gaps due to changing network delay, immediately releasing blocks upon recovery exposes this variability to the application.

In many practical workloads, data entering LT3 naturally exhibits burst structure that reflects higher-level organization, such as video frames or inference outputs. As a result, the block partitioning used for transmission often aligns with this structure, even though LT3 itself does not depend on application semantics. Preserving the timing and ordering of these reconstructed blocks therefore indirectly preserves the behavior expected by the application.

Sender-side techniques help but do not remove the problem. Pacing can smooth departures when path conditions are stable, but it struggles to track changes in path delay over time and cannot correct variability introduced by queueing, reordering, or recovery dynamics that only the receiver observes. Larger playback buffers can hide jitter, yet they increase end-to-end latency and reduce responsiveness. Traditional jitter buffers use fixed or slowly adaptive depth rules that are poorly matched to the variability patterns seen in modern low-latency systems, especially when data is reconstructed from multiple packets.

This report introduces Adaptive Release Control (\arc), a receiver-side release scheduling protocol that smooths delivery timing by operating at the point where recovery-time variability becomes visible. \arc\ uses timestamps generated by the LT3 sender together with local recovery times to release data blocks in a manner that reflects the sender’s timing, thereby maintaining ordering and limiting reordering when necessary, and producing stable release timing with minimal additional latency given network conditions. It operates entirely on the receiver clock and requires no feedback, synchronization, or changes to the underlying transport.

\arc\ includes additional mechanisms useful in practical systems. Quantized scheduling locks release times to increments aligned with the timing of block transmissions at the LT3 sender and suppresses micro-level jitter. A bounded in-order guard preserves sequencing when recovery completes out of order while preventing unbounded waiting for delayed predecessors. These refinements rely only on local timing information and do not alter the sender or underlying transport.

The approach integrates cleanly with LT3, a network-layer system currently deployed as an overlay \cite{AggarwalLubyMinder2025Immersive}. LT3 operates between cooperating endpoints and forwards packets without altering the transport protocols it carries. In typical deployments, the LT3 sender and receiver are placed close to the application endpoints, so most timing distortion arises on the path between them. Receiver-side scheduling restores this timing at the LT3 receiver before reconstructed blocks are released from LT3 toward the application.

\arc\ applies to traffic carried through LT3 independent of the specific protocols or data formats involved. When traffic such as TCP, QUIC, WebRTC, RTP, or UDP is carried through LT3 with \arc, preserving the timing of data entering the LT3 sender at the receiver improves behavior across a wide range of conditions. This effect is demonstrated in the evaluation sections below.

\paragraph{Contributions.}
\begin{itemize}
  \item A receiver-side scheduling approach that restores sender timing by adapting release timing to the observed recovery pattern, including asymmetric offset adjustment, quantized scheduling, and bounded in-order release.
  \item An integration of this approach into the LT3 system, which shows that receiver-side timing control can be added as a modular component without modifying the application's transport protocol.
  \item A real-world evaluation of LT3 with receiver-side scheduling, using a cloud-gaming workload, that shows substantial reductions in jitter, significantly tighter delivery intervals, and materially improved percentile latency consistency.
\end{itemize}

\section{Related Work}
Receiver-side buffering and playout control have been widely studied across real-time media and streaming systems.

Early adaptive playout protocols for packet audio estimated network delay and dynamically adjusted the receiver playout point to smooth jitter \cite{ramjee1994}. Later work integrated forward error correction with delay adaptation \cite{rose2000} and refined jitter-buffer control for VoIP environments to balance delay against late-loss probability \cite{maity2018,bhunia2011}. These designs typically reset or retune delay during talkspurts and assume independent, small packets rather than block recovery.

Within the RTP/RTCP framework, inter-arrival jitter measurement and receiver-side buffering are standardized in the RTP specification \cite{rfc3550} and further refined for transmission-time offsets \cite{rfc5450}. Most RTP implementations maintain a fixed or slowly adaptive jitter buffer whose depth is chosen heuristically. In practice this means using a preset value, often 50--100~ms, or a simple moving-average rule such as
\[
\text{buffer} = \text{mean delay} + k \times \text{std deviation},
\]
with \(k\) empirically selected, commonly between 1 and 2. Some systems enlarge the buffer after several late packets and then shrink it gradually when conditions stabilize, while others ship codec-specific presets determined through laboratory testing and subjective playback evaluation. These methods rely on heuristics tuned by measurement and experience rather than on an explicit control model or analytic optimization.

Industrial low-latency streaming protocols expose similar receiver buffer parameters. The Secure Reliable Transport (SRT) protocol defines a latency configuration that directly sets the receiver buffer depth and consequently the minimum end-to-end delay \cite{srt-doc-latency,srt-doc-hmg}. The IETF Internet-Draft for SRT describes buffering behavior but does not specify a receiver-side scheduling protocol beyond this adjustable delay window \cite{srt-draft}. In practice, SRT and related contribution protocols rely on static configuration and are sensitive to network variability.

In HTTP adaptive streaming, buffer management is primarily used for rate adaptation rather than post-recovery scheduling. Buffer-based controllers select segment bitrates based on playback buffer occupancy \cite{huang2014}, and work such as FESTIVE analyzes the fairness and stability of such adaptive bitrate protocols \cite{festive2012}. These mechanisms regulate transmission rate and encoding level rather than inter-release timing at the receiver.

The present work differs in focusing on scheduling after recovery, at the point where variability becomes visible to the receiver. Unlike prior approaches that adjust buffering or playout delay, \arc\ directly regulates the timing at which recovered data is released to the application. This introduces a receiver-side scheduling layer between recovery and application delivery that operates independently of transport behavior and sender pacing.
\section{Model and Problem Statement}
Although we describe the protocol in terms of LT3 blocks, the same mechanisms apply more generally to any sequence of data units reconstructed at the receiver.
We consider a sender that generates blocks indexed by \(n\) with LT3 sender timestamp \(S_n\) and a receiver that recovers them at times \(A_n\) on its local clock. We use “sender” and “receiver” to denote the ingress and egress points of the LT3 system, that is, where packets enter LT3 at the sender and where reconstructed data is released at the receiver. The goal is to preserve the timing of data entering LT3 at the sender as it is released from LT3 at the receiver. The task is to choose release times \(T_n\) that satisfy three goals:
\begin{enumerate}[label=(\roman*)]
  \item preserve order or permit only bounded overtake,
  \item keep the transmission-to-release delay small and controlled,
  \item match the inter-release intervals to the timing between block formation and transmission at the LT3 sender as closely as possible.
\end{enumerate}

Clocks are not synchronized. Receiver-side scheduling uses LT3 sender timestamps as metadata and otherwise operates entirely on the receiver clock. In LT3, these timestamps are generated at the sender based on packet arrival times, so preserving this timing at the receiver preserves the arrival timing at the LT3 sender, supporting smooth delivery and stable buffering.

\paragraph{Design goals.}
We target four concrete properties that guide the controller:
\begin{enumerate}[label=(G\arabic*)]
  \item one-to-one correspondence between sent and delivered blocks,
  \item preservation of the original order except for bounded, controlled overtake,
  \item equality of inter-block spacing at sender and receiver up to small bounded error,
  \item minimal additional latency beyond what is needed for smooth delivery.
\end{enumerate}

\section{Adaptive Release Control Protocol}
This section formalizes the Adaptive Release Control (\arc) protocol. The receiver maintains an offset \(D\) that maps LT3 sender timestamps onto the receiver's time axis. For any block with LT3 sender timestamp \(S_n\), the expression \(S_n + D\) represents a projected release time on the receiver clock. The scheduler then compares this projected value against the recovery time \(A_n\) to determine when the block should be released. The offset \(D\) is therefore a continuously adjusted translation between the sender and receiver clocks, chosen so that the projected release time \(S_n + D\) lies at or just beyond the recovery time for most blocks. The release time is always at least the larger of the recovery time and the projected value, and it keeps the scheduled times close to the upper envelope of the observed recovery times. The controller updates \(D\) asymmetrically, reacting quickly to late arrivals while reducing the offset gradually when arrivals are early.

\subsection{State, Variables, and Initialization}
Let \(D\) denote the adaptive mapping offset. For the first recovered block with \((S_0, A_0)\), initialize
\[
D \leftarrow A_0 - S_0,
\]
so that the initial projected release time \(S_0 + D\) coincides with its recovery time \(A_0\).

To prevent drift during idle periods, if no recoveries occur for longer than a timeout \(T_{\mathrm{idle}}\), the controller re-anchors the offset when activity resumes by setting
\[
D \leftarrow A_n - S_n
\]
for the first block \(n\) recovered after the idle gap. This reinitialization restores alignment between LT3 sender timestamps and the receiver’s clock after a prolonged pause. The controller does not reinitialize for any block that is recovered within \(T_{\mathrm{idle}}\) of its predecessor.

For each subsequent recovered block \(n\), define the deviation
\[
X_n = A_n - (S_n + D),
\]
which is the difference, in receiver time, between the recovery time and the projected release time implied by the current offset. Positive deviations indicate that the block recovered later than the projected schedule; negative deviations indicate recovery earlier than projected.

\subsection{Asymmetric Update}
To stabilize scheduling while preserving responsiveness to delay spikes, \arc\ applies an asymmetric update to \(D\). The update uses exponent parameters \(0 \le \rho_u \le 1 \le \rho_\ell\), damping parameters \(\lambda_u > 0\) and \(\lambda_\ell > 0\), and a clipping parameter \(U>0\). A typical choice for these parameters is \(\rho_u = 0.5\), \(\rho_\ell = 1.5\), \(\lambda_u = 0.05\), \(\lambda_\ell = 0.03\), and \(U = 1\) second.

Given \(X_n\), \arc\ computes an adjustment \(Y_n\) and then updates \(D\) as follows:
\[
\text{if } X_n > 0:\quad
Y_n = \left(\frac{\min\{X_n, U\}}{U}\right)^{\rho_u} \cdot U,\qquad
D \leftarrow D + \lambda_u \cdot Y_n,
\]
\[
\text{if } X_n < 0:\quad
Y_n = -\left(\frac{\min\{|X_n|, U\}}{U}\right)^{\rho_\ell} \cdot U,\qquad
D \leftarrow D + \lambda_\ell \cdot Y_n.
\]

The exponent parameters are chosen so that the protocol reacts more strongly to late recoveries than to early ones. This effect arises primarily from the asymmetric shaping of \(Y_n\) by \(\rho_u\) and \(\rho_\ell\):

\begin{itemize}
\item \textbf{Exponent asymmetry.}
Since \(0 \le \rho_u \le 1\), the function \(a \mapsto a^{\rho_u}\) is concave on \([0,U]\). For any late deviation \(X_n = a \in (0,U]\),
\[
Y_n = \left(\frac{a}{U}\right)^{\rho_u} \cdot U \ge a,
\]
so the change in \(D\) overshoots the observed lateness before damping by \(\lambda_u\). This overshoot allows the projected release timeline to move above current delay peaks.

Since \(\rho_\ell \ge 1\), the function \(a \mapsto a^{\rho_\ell}\) is convex on \([0,U]\). For any early deviation \(X_n = -a\) with \(a \in (0,U]\),
\[
|Y_n| = \left(\frac{a}{U}\right)^{\rho_\ell} \cdot U \le a,
\]
so the change in \(D\) undershoots the amount by which recovery is early. This yields gentler downward motion into delay valleys.

\item \textbf{Damping parameters.}
The damping parameters \(\lambda_u\) and \(\lambda_\ell\) scale these overshoot and undershoot adjustments before they are added to \(D\). They control how aggressively the offset responds to the shaped deviations and can be chosen independently. In practice, choosing \(\lambda_u\) and \(\lambda_\ell\) on the order of a few percent of \(Y_n\) produces gradual but responsive adaptation.
\end{itemize}

Together, the concave late-response shaping and convex early-response shaping, combined with damping, shift the projected release timeline forward rapidly when recovery lags (\(X_n>0\)), reducing the risk of late release, while allowing it to drift downward more cautiously when recovery is early. This keeps the projected release times aligned with the delay peaks rather than the midpoint of recovery variability.

We optionally apply a bound \(\delta>0\) on waiting through
\[
D \leftarrow \min\bigl(D,\, A_n - S_n + \delta\bigr),
\]
which limits the projected release time \(S_n + D\) to exceed the recovery time \(A_n\) by at most \(\delta\).

\paragraph{Why asymmetry helps.}
Recovery patterns naturally contain peaks and valleys. If reactions to early and
late recoveries were symmetric, the offset \(D\) would track the midpoint of
these fluctuations and would routinely fall below the true delay peaks, which
increases the risk of late release and playback stalls. As shown in
Section~\ref{sec:analytic}, the symmetric linear case
\(\rho_u = \rho_\ell = 1\) cannot maintain the projected timeline at or above
the upper envelope of a fluctuating recovery pattern; instead it converges to a
value strictly between the high and low delays. The exponent asymmetry described
above directly addresses this limitation.

When \(X_n > 0\) is small (block recovered slightly late), the concave shaping
with \(\rho_u < 1\) produces an adjustment \(Y_n\) whose magnitude exceeds
\(X_n\) before damping, so the projected release timeline is pushed upward by
more than the observed lateness. This overshoot makes it easier for subsequent
blocks to recover before their scheduled release. When \(X_n < 0\) is small
(block recovered slightly early), the convex shaping with \(\rho_\ell > 1\)
produces an adjustment whose magnitude is less than \(|X_n|\), so the projected
timeline moves downward only slightly. The combination stays near the upper
envelope of the recovery pattern and avoids deep excursions into delay valleys.

A further benefit comes from the stability of inter-release spacing. When the
projected release time of a block is after its recovery time, the release
interval matches the original inter-send interval provided that \(D\) remains
nearly constant. Because the asymmetric update rules make early recoveries more common than late ones, and the early adjustments are undershooting, \(D\) changes very little from one block to the next even under varying network conditions.
As a result,
the inter-release times at the receiver stay close to the inter-send times at the
sender, which supports smooth playback.

\paragraph{Optional neutral band.}
An optional neutral band parameter \(J \ge 0\) can be used to suppress updates for modest early recoveries and to retain a small safety margin in the release rule. When \(J>0\) is enabled, the early-update condition and adjustment are modified to
\[
\text{if } X_n \le -J:\quad
Y_n = -\left(\frac{\min\{\bigl|X_n + J\bigr|, U\}}{U}\right)^{\rho_\ell} \cdot U,\qquad
D \leftarrow D + \lambda_\ell \cdot Y_n,
\]
so that there is no update when \(X_n \in (-J,0]\). In this regime, the projected release time is already close enough to the recovery time that additional downward adjustment is unnecessary. In some deployments and in the evaluation below, \(J\) is set to zero and the core \arc\ behavior described above dominates. The neutral band parameter \(J\), when used, suppresses small downward corrections and preserves this behavior while absorbing noise.

\subsection{Release Scheduling}
After updating \(D\), the scheduler assigns the release time. In the core \arc\ form without a neutral band,
\[
T_n = \max\{A_n,\, S_n + D\}.
\]
All quantities are expressed on the receiver clock. This rule ensures that:
\begin{itemize}
    \item a block is scheduled to be released after it has been recovered (\(T_n \ge A_n\)),
    \item the release timing follows the timing at the LT3 sender through the projected mapping \(S_n + D\),
    \item inter-release spacing remains close to the inter-send spacing as long as \(D\) varies slowly.
\end{itemize}

When a neutral band \(J>0\) is enabled, the scheduler includes a small safety margin and uses
\[
T_n = \max\{A_n,\, S_n + D + J\},
\]
which places most recoveries shortly before their release times and maintains the envelope-tracking behavior described above.

\section{Illustrative Behavior}
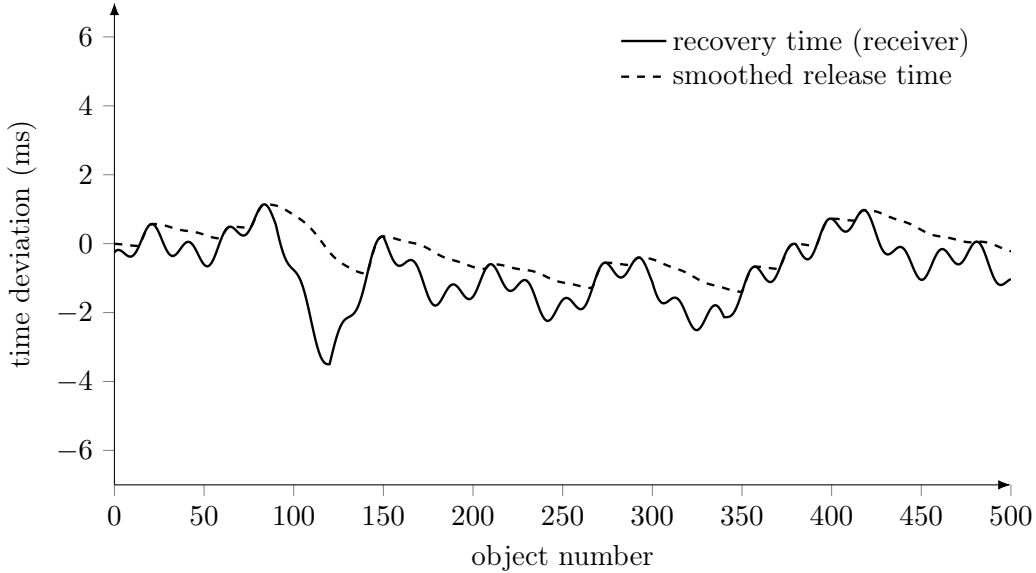
\begin{figure}[H]
\centering
\begin{tikzpicture}
\begin{axis}[
  width=13.5cm, height=8.0cm,
  axis lines=left, axis line style={-Latex},
  xlabel={block number}, ylabel={time deviation (ms)},
  xmin=0, xmax=500, ymin=-7, ymax=7,
  tick align=outside, ticks=both,
  legend style={draw=none, fill=none},
  legend pos=north east,
  legend cell align=left,
  unbounded coords=discard,
  every axis plot/.style={line width=0.9pt},
]
  \pgfmathsetmacro{\attack}{0.85}
  \pgfmathsetmacro{\release}{0.02}
  \def\rawcoords{}
  \def\smoothcoords{}
  \pgfmathsetmacro{\sPrev}{0.0}
  \foreach \k in {0,...,500} {
    \pgfmathsetmacro{\dipA}{max(0, 1 - abs(\k-120)/30)}
    \pgfmathsetmacro{\dipB}{max(0, 1 - abs(\k-340)/40)}
    \pgfmathsetmacro{\rk}{
      -0.6
      + 0.9*sin(deg(0.020*\k))
      + 0.5*sin(deg(0.095*\k))
      + 0.35*cos(deg(0.300*\k))
      - 3.0*\dipA
      - 2.4*\dipB
    }
    \pgfmathsetmacro{\sk}{
      (\rk > \sPrev) * (\sPrev + \attack*(\rk - \sPrev))
      + (\rk <= \sPrev) * (\sPrev - \release*(\sPrev - \rk))
    }
    \xdef\rawcoords{\rawcoords (\k,\rk)}
    \xdef\smoothcoords{\smoothcoords (\k,\sk)}
    \xdef\sPrev{\sk}
  }
  \addplot[black, solid] coordinates {\rawcoords};
  \addlegendentry{recovery time (receiver)}
  \addplot[black, dashed] coordinates {\smoothcoords};
  \addlegendentry{smoothed release time}
\end{axis}
\end{tikzpicture}
\caption{Recovery times versus smoothed release times.}
\label{fig:smoothing-example}
\end{figure}

Figure~\ref{fig:smoothing-example} compares raw recovery timing against the smoothed schedule generated by \arc. The envelope-following behavior is visible: when recovery rises above the schedule, the schedule rises quickly; in flat regions it decays only slowly, which limits added delay while keeping most arrivals ahead of their release deadlines.

\section{System View}
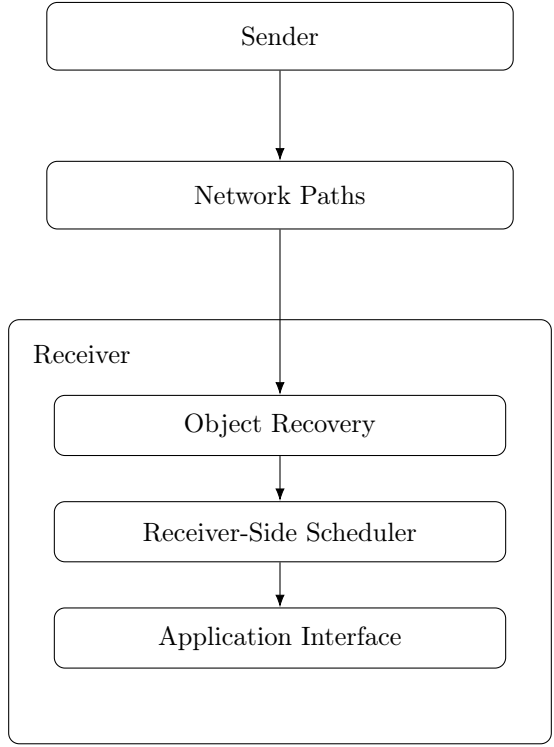
\begin{figure}[H]
\centering
\begin{tikzpicture}[
  font=\small, node distance=10mm, >=Latex,
  box/.style={draw, rounded corners, minimum width=6.2cm, minimum height=9mm, align=center},
  sub/.style={draw, rounded corners, minimum width=6.0cm, minimum height=8mm, align=center, fill=white}
]
  \node[box]                           (sender)  {Sender};
  \node[box, below=12mm of sender]     (network) {Network Paths};
  \node[sub, below=22mm of network]    (decoder)  {Block Recovery};
  \node[sub, below=6mm  of decoder]    (smoother) {Receiver-Side Scheduler};
  \node[sub, below=6mm  of smoother]   (appif)    {Application Interface};

  \begin{scope}[on background layer]
    \node[draw, rounded corners, inner ysep=10mm, inner xsep=6mm,
          fit=(decoder)(smoother)(appif)] (receiver) {};
  \end{scope}
  \node[anchor=north west, xshift=2mm, yshift=-2mm] at (receiver.north west) {Receiver};

  \draw[->] (sender)  -- (network);
  \draw[->] (network) -- (decoder);
  \draw[->] (decoder) -- (smoother);
  \draw[->] (smoother) -- (appif);
\end{tikzpicture}
\caption{Receiver-side scheduling sits after recovery and before release to the application.}
\label{fig:system-overview}
\end{figure}

Figure~\ref{fig:system-overview} shows where receiver-side scheduling fits in the receive pipeline. Packets from one or more network paths feed into a recovery module that reconstructs each block and produces its LT3 sender timestamp \(S_n\) together with its recovery time \(A_n\). The scheduling module runs immediately after recovery and uses these \((S_n, A_n)\) pairs to determine a release time \(T_n\) for each block. This placement ensures that the scheduler operates on the timing actually experienced by the receiver while remaining independent of the underlying transport protocol.

Because it observes only LT3 sender timestamps and local recovery times, the same module can follow TCP reassembly, QUIC datagram delivery, WebRTC frame construction, RTP depacketization, or block-based coded systems such as LT3. The scheduler neither modifies transport behavior nor requires feedback to the sender; it simply regulates when recovered blocks are presented to the application interface. This separation allows receiver-side scheduling to be added as a lightweight component in existing media and control pipelines.

\section{Enhanced \arc}

Receiver-side scheduling in its basic \arc\ form already stabilizes release timing by maintaining an adaptive offset that tracks the upper envelope of observed recovery times. In practical deployments, however, several additional refinements improve timing stability, sequencing behavior, and responsiveness to real network dynamics. These refinements operate entirely on the receiver clock, require no sender changes, and preserve the core control logic of \arc.

This section introduces three enhancements that build on the adaptive offset mechanism. Quantized scheduling regularizes small fluctuations by following the adaptive offset through discrete steps that align with the sender’s timing. The bounded in-order guard preserves sequencing when recovery completes out of order while avoiding unbounded head-of-line waiting. Finally, automatic parameter adaptation allows envelope scales, clamp limits, and guard windows to track prevailing RTT and recovery characteristics so that the scheduler remains well tuned across a wide range of path and workload conditions.

Each enhancement is optional and independent. Together they provide a practical extension of the basic \arc\ mechanism suitable for real-time applications such as cloud gaming, XR streaming, and interactive media pipelines.

\label{sec:enhanced_ADC}

\subsection{Quantized Steps}
\arc\ removes coarse jitter but may leave small block-to-block variations in \(D\). 
\arc\ with quantization (\arcq) introduces a quantized offset \(E\) that follows \(D\) in steps of size \(\gamma > 0\), where \(\gamma\) is a configurable quantization step parameter. The result is a perceptually smooth timing with bounded tracking error.

\paragraph{Midpoint hysteresis.}
Initialize
\[
D \leftarrow A_0 - S_0,\qquad E \leftarrow D + \frac{\gamma}{2}.
\]
Hold \(E\) while \(D \in [E-\gamma, E]\). If \(D\) exits the band, re-anchor
\[
E \leftarrow D + \frac{\gamma}{2},
\]
which preserves \(E \ge D\) and \(0 \le E - D < \gamma\). Release with
\[
T_n = \max\{A_n,\, S_n + E + J\}.
\]
Choosing \(\gamma\) on the order of the spacing of data arrivals at the sender, such as a video frame interval, suppresses micro-stutter while keeping \arcq\ agile.

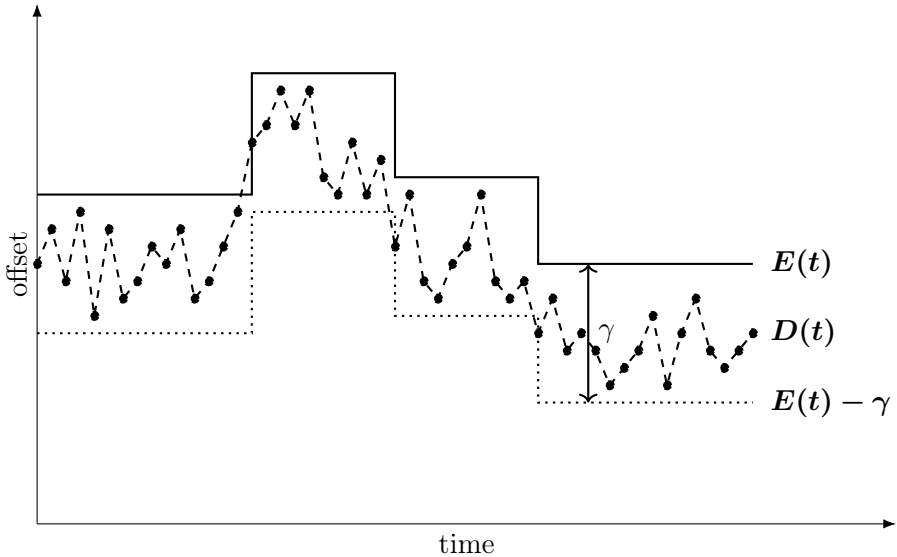
\begin{figure}[H]
\centering
\begin{tikzpicture}
\begin{axis}[
  width=13cm, height=8.5cm,
  axis lines=left, axis line style={-Latex},
  xlabel={time}, ylabel={offset},
  xmin=0, xmax=12, ymin=0.0, ymax=3.0,
  ticks=none, clip=false,
  every axis plot/.style={line width=0.9pt},
]
  \def\gstep{0.8}
  \addplot+[name path = Dtime, color=black, mark=*,mark size = 1.5pt,
          mark options={fill=black, draw=black}, thick, dashed] 
  coordinates {
    (0.0, 1.5) (0.2, 1.70)
    (0.4, 1.4) (0.6, 1.80)
    (0.8, 1.2) 
    (1.0, 1.7) (1.2, 1.30)
    (1.4, 1.4) (1.6, 1.60)
    (1.8, 1.5) 
    (2.0, 1.7) (2.2, 1.30)
    (2.4, 1.4) (2.6, 1.60)
    (2.8, 1.8) 
    (3.0, 2.2) (3.2, 2.30)
    (3.4, 2.5) (3.6, 2.30)
    (3.8, 2.5) 
    (4.0, 2.0) (4.2, 1.9)
    (4.4, 2.2) (4.6, 1.9)
    (4.8, 2.1) 
    (5.0, 1.6) (5.2, 1.9)
    (5.4, 1.4) (5.6, 1.3)
    (5.8, 1.5) 
    (6.0, 1.6) (6.2, 1.9)
    (6.4, 1.4) (6.6, 1.3)
    (6.8, 1.4) 
    (7.0, 1.1) (7.2, 1.3)
    (7.4, 1.0) (7.6, 1.1)
    (7.8, 1.0) 
    (8.0, 0.8) (8.2, 0.9)
    (8.4, 1.0) (8.6, 1.2)
    (8.8, 0.8) 
    (9.0, 1.1) (9.2, 1.3)
    (9.4, 1.0) (9.6, 0.9)
    (9.8, 1.0) 
    (10.0, 1.1)
    };
  \addplot+[name path=Etop, const plot, thick, color=black, mark=none]
    coordinates {
      (0.0, 1.9) (3.0, 1.9)
      (3.0, 2.6) (5.0, 2.6)
      (5.0, 2.0) (7.0, 2.0)
      (7.0, 1.5) (10.0, 1.5)
    };
  \addplot+[name path=Ebot, dotted, color=black, mark=none, thick, const plot]
    coordinates {
      (0.0, 1.9-\gstep) (3.0, 1.9-\gstep)
      (3.0, 2.6-\gstep) (5.0, 2.6-\gstep)
      (5.0, 2.0-\gstep) (7.0, 2.0-\gstep)
      (7.0, 1.5-\gstep) (10.0, 1.5-\gstep)
    };
  \pgfmathsetmacro{\xlab}{10.2}
  \pgfmathsetmacro{\Dlab}{1.1}
  \pgfmathsetmacro{\Elab}{1.5}
  \node[fill=white, inner sep=1pt, anchor=west] at (axis cs:\xlab,\Dlab) {\(\bm{D(t)}\)};
  \node[fill=white, inner sep=1pt, anchor=west] at (axis cs:\xlab,\Elab) {\(\bm{E(t)}\)};
  \node[fill=white, inner sep=1pt, anchor=west] at (axis cs:\xlab,\Elab-\gstep) {\(\bm{E(t)-\gamma}\)};
  \pgfmathsetmacro{\xg}{7.7}
  \pgfmathsetmacro{\Eg}{\Elab}
  \draw[<->, thick] (axis cs:\xg,\Eg) -- (axis cs:\xg,\Eg-\gstep);
  \node[fill=white, inner sep=1pt, anchor=west] at (axis cs:\xg+0.1,\Eg-0.5*\gstep) {\(\gamma\)};
\end{axis}
\end{tikzpicture}
\caption{\arcq\ scheduling with midpoint hysteresis.}
\label{fig:quantization}
\end{figure}

\paragraph{Properties.}
The invariants \(E \ge D\) and \(0 \le E - D < \gamma\) bound added latency and enforce discrete, predictable adjustments. Re-anchoring requires a finite movement in \(D\), which prevents rapid toggling due to noise. Because re-anchoring sets \(E \leftarrow D + \gamma/2\), the quantized offset \(E\) changes only when \(D\) moves by at least \(\gamma/2\). Small fluctuations in \(D\) therefore leave \(E\) unchanged, so the inter-release timing shifts only when the underlying offset changes by a meaningful amount. Figure~\ref{fig:quantization} illustrates how \(E(t)\) tracks the smoothed offset \(D(t)\) with a hysteresis band of width \(\gamma\) and only re-anchors when \(D(t)\) crosses the band boundary.

\subsection{Bounded In-Order Guard}
Some transports and applications assume ordered delivery. \arcq\ with guard (\arcqg) adds a guard interval \(G \ge 0\), a configurable guard-window parameter, to honor this assumption without creating unbounded head-of-line blocking.

\begin{figure}[H]
\centering
\begin{tikzpicture}[x=1cm,y=1cm,>=Latex,font=\small]
  \draw[-Latex] (0,0) -- (11,0) node[anchor=west]{time};
  \node[anchor=east] at (0,0.45) {block \(n\)};
  \draw[densely dashed] (3,0.45) -- (3,0);
  \node[above] at (3,0.45) {\(T_n\)};
  \draw[densely dashed] (9,0.45) -- (9,0);
  \node[above] at (9,0.45) {\(T_n + G\)};
  \draw[decorate,decoration={brace, mirror, amplitude=4pt}] (3,-0.35) -- (9,-0.35)
    node[midway, yshift=-12pt] {\(G\)};
\end{tikzpicture}
\caption{Bounded in-order guard window.}
\label{fig:guard-timing}
\end{figure}
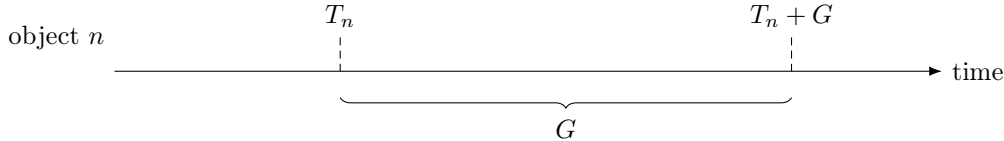

\paragraph{Rule.}
With candidate \(T_n\) from \arc\ or \arcq:
\begin{itemize}
  \item if \(n-1\) is released by \(T_n\), release \(n\) at \(T_n\),
  \item if \(n-1\) releases in \([T_n,\, T_n+G)\), release \(n\) immediately after \(n-1\),
  \item otherwise release \(n\) at \(T_n+G\).
\end{itemize}
This rule preserves order when feasible while bounding additional waiting if a predecessor is delayed. Selecting \(G\) near a high percentile of observed reordering covers most cases without material latency cost. Figure~\ref{fig:guard-timing} shows the guard window of length \(G\) around \(T_n\), indicating when block \(n\) waits for its predecessor and when it releases at \(T_n + G\) if the predecessor remains missing.

\subsection{Automatic Parameter Adaptation}

Several parameters in \arc, \arcq, and \arcqg\ reflect the scale of variability
seen at the receiver and need not remain fixed. In LT3, the receiver has access
to auxiliary timing measurements such as RTT that are maintained within the LT3
system itself. These measurements operate on the receiver clock and provide a
meaningful physical time scale, unlike the offset \(D\), whose value depends on
unrelated sender–receiver clock alignment.

This section describes how envelope, clamp, and guard parameters can be driven
by RTT and by statistics of \emph{inter-recovery intervals}, which are fully
defined on the receiver clock and do not rely on LT3 sender timestamps.

\paragraph{Envelope scale from RTT.}
The clipping parameter \(U\) governs how aggressively the protocol responds to
large bursts in recovery timing. Because \(U\) represents a real time duration,
it should be tied to a physically meaningful baseline such as RTT.

Let \(\mathrm{RTT}_{\mathrm{sm}}\) be a smoothed RTT estimate supplied by LT3. 
A robust envelope scale is then
\[
U \approx c_U \cdot \mathrm{RTT}_{\mathrm{sm}},
\]
where \(c_U \in [0.5, 2.0]\) depends on the desired responsiveness. This keeps
the response window aligned with real path dynamics and avoids interpreting
\(D\) or \(A_n - S_n\) as physical delays.

\paragraph{Clamp parameter from observed variability.}
The clamp parameter \(\delta\) limits how far the projected release time
\(S_n + D + J\) may exceed the recovery time \(A_n\). Since \(D\) is an abstract
offset, \(\delta\) must depend only on quantities that are time-meaningful at
the receiver.

Let \(\Delta_n = A_n - A_{n-1}\) denote inter-recovery spacings on the receiver
clock, and let \(q_{p}\) be the empirical \(p\)-th percentile over a sliding
window. A practical clamp choice is
\[
\delta \approx c_\delta \cdot (q_{95} - q_{50}),
\]
which scales allowable additional waiting to the spread between median and
tail inter-recovery intervals. Updating this estimate slowly (e.g., once per
several hundred blocks) prevents the clamp from reacting to short-lived spikes.

\paragraph{Adaptive guard window from reordering statistics.}
The guard parameter \(G\) in \arcqg\ bounds waiting for predecessor blocks when
recovery occurs out of order. Reordering is reflected directly in recovery-time
differences.

Define the skew
\[
R_n = A_n - A_{n-1},
\]
and treat \(R_n < 0\) as reordering events. Let
\(q^{\mathrm{reorder}}_{p}\) be the empirical \(p\)-th percentile of \(|R_n|\)
for these events. A suitable guard window is
\[
G \approx q^{\mathrm{reorder}}_{p},
\]
for example using \(p = 95\), which covers most reorderings while bounding
worst-case waiting.

\paragraph{Interaction with \arc\ shaping parameters.}
The adaptive choices for \(U\), \(\delta\), and \(G\) are orthogonal to the
shaping parameters \((\rho_u,\rho_\ell)\) and damping parameters
\((\lambda_u,\lambda_\ell)\). The exponent parameters determine overshoot and
undershoot behavior, while the adaptive parameters set the time scale on which
the behavior operates.

This separation enables a practical deployment strategy:
\begin{itemize}
\item tune \((\rho_u,\rho_\ell,\lambda_u,\lambda_\ell)\) once per system,
\item adapt \((U,\delta,G)\) automatically from RTT and inter-recovery
      statistics at each receiver.
\end{itemize}

\paragraph{Influence of application-level constraints.}
In addition to RTT and recovery dynamics, systems may expose application-level
signals that influence desirable operating points. Examples include latency
budgets, frame deadlines, control-loop sensitivity, or other application
requirements that define acceptable waiting or variability. These signals may
be incorporated into the adaptation logic to bias parameters toward more
aggressive or more conservative smoothing.

Such influences are intentionally kept abstract here: the \arc\ mechanisms
operate independently of any specific application semantics, and the adaptation
rules above function without assuming any particular higher-level policy.
However, where application constraints exist, they can guide the automatic
selection of \(U\), \(\delta\), and \(G\) to balance responsiveness and
latency in a manner consistent with the needs of the deployment.

\section{Protocols}
We present concise pseudocode for \arc, \arcq, and \arcqg. Parameters and symbols follow the definitions above. The listings describe per-block update and release rules applied at the receiver, and are intended to illustrate the scheduling logic independent of implementation details.

\subsection*{\arc: Adaptive Offset}
\begin{algorithm}[H]
\caption{\arc: adaptive offset update and scheduling}
\begin{algorithmic}[1]
\Require damping parameters \(\lambda_u,\lambda_\ell > 0\); exponents \(\rho_u,\rho_\ell \ge 0\);
clipping \(U>0\); neutral band \(J\ge0\); max delay \(\delta>0\); idle timeout \(T_{\mathrm{idle}}>0\)
\Statex \textit{State:} offset \(D\); last recovery time \(\tau_{\mathrm{last}}\)
\State Initialize \(D \gets A_0 - S_0,\ \tau_{\mathrm{last}} \gets A_0\)
\For{each recovered block with \((S_n,A_n)\)}
  \If{\(A_n - \tau_{\mathrm{last}} \ge T_{\mathrm{idle}}\)} \Comment{idle re-anchor}
    \State \(D \gets A_n - S_n\)
  \EndIf
  \State \(\tau_{\mathrm{last}} \gets A_n\); \(X \gets A_n - (S_n + D)\)
  \If{\(X > 0\)}
    \State \(Y \gets \left(\frac{\min\{X,U\}}{U}\right)^{\rho_u} \cdot U\); \(D \gets D + \lambda_u \cdot Y\)
  \ElsIf{\(X \le -J\)}
    \State \(Y \gets -\left(\frac{\min\{|X+J|,U\}}{U}\right)^{\rho_\ell} \cdot U\); \(D \gets D + \lambda_\ell \cdot Y\)
  \EndIf
  \State \(D \gets \min(D,\, A_n - S_n + \delta)\)
  \State \(T_n \gets \max\{A_n,\, S_n + D + J\}\) \Comment{schedule}
  \State enqueue for release at \(T_n\)
\EndFor
\end{algorithmic}
\end{algorithm}

\subsection*{\arcq: Quantized Offset}
\begin{algorithm}[H]
\caption{\arcq: quantized scheduling with midpoint hysteresis}
\begin{algorithmic}[1]
\Require parameters of \arc\ plus step \(\gamma>0\)
\State Initialize \(D \gets A_0 - S_0\), \(E \gets D + \gamma/2\)
\For{each recovered block}
  \State update \(D\) as in \arc
  \If{\(D > E\) \textbf{ or } \(D < E - \gamma\)}
    \State \(E \gets D + \gamma/2\)
  \EndIf
  \State \(T_n \gets \max\{A_n,\, S_n + E + J\}\); enqueue for release at \(T_n\)
\EndFor
\end{algorithmic}
\end{algorithm}

\subsection*{\arcqg: Quantized with Bounded Guard}
\begin{algorithm}[H]
\caption{\arcqg: ordered release with guard \(G\)}
\begin{algorithmic}[1]
\Require parameters of \arcq\ plus guard \(G \ge 0\)
\For{each recovered block \(n\)}
  \State compute \(T_n\) from \arcq
  \If{\(n-1\) released by \(T_n\)} release \(n\) at \(T_n\)
  \ElsIf{\(n-1\) releases in \([T_n,\,T_n+G)\)} release \(n\) immediately after \(n-1\)
  \Else\ release \(n\) at \(T_n + G\)
  \EndIf
\EndFor
\end{algorithmic}
\end{algorithm}


\section{Implementation Notes}

The scheduler operates on reconstructed data blocks produced by LT3, using recovery events and LT3 sender timestamps to determine release timing. It is independent of the nature of the traffic carried through LT3 and applies equally when the carried traffic consists of packets, multiple concurrent flows, or other data formats. Per-block work consists of a small number of arithmetic operations. In practice it is useful to exclude small control blocks from smoothing via a minimum size threshold and release them immediately.

\paragraph{Representative static settings.}
\[
\rho_u \in [0.3,0.7],\quad \rho_\ell \in [1.0,2.0],
\quad \lambda_u \in [0.10,1.0],\quad \lambda_\ell \in [0.01,0.10],
\]
\[
U \in [50,200]\ \text{ms},
\quad
\delta \in [30,100]\ \text{ms},\quad
\gamma \in [8,20]\ \text{ms},
\]
\[
T_{\mathrm{idle}} \in [0.5,2.0]\ \text{s},
\quad
 J \in [0,5]\ \text{ms},\quad
G \in [20,200]\ \text{ms}.
\]


\section{Integration of \arcq\ into LT3}
\label{sec:LT3_integration}

The receiver-side scheduling mechanisms in this report are implemented within LT3 \cite{AggarwalLubyMinder2025Immersive}, a network-layer system that operates across heterogeneous paths and can employ deterministic spraying for robustness \cite{LubyByers2025Whack}. LT3 transmits data as forward-error-protected blocks and reconstructs each block at the receiver once sufficient coded packets have arrived. Receiver-side scheduling then determines when each recovered block is released toward the application using the \arcq\ protocol.

\paragraph{Placement in the receive path.}
Coded packets are generated at the sender, forwarded through LT3, and decoded at the receiver into complete blocks. The receiver-side scheduling module runs immediately after decoding and before the application interface. Each recovered block carries its LT3 sender timestamp \(S_n\) and recovery time \(A_n\); the controller computes a release time \(T_n\) and releases the block when the receiver clock reaches \(T_n\). This isolates timing control from lower-layer variability and bases decisions on observed recovery timing.

\paragraph{Operational integration and benefits.}
Decoding completion times reflect combined effects of path latency, congestion, and coding overhead. Immediate release produces irregular inter-arrival intervals even when throughput is stable. \arcq\ smooths these completion events without additional coordination, maintaining stable timing for playback or upstream pipelines. The module is independent of the application's transport protocol and adds constant-time work per recovered block.


\section{Evaluation}

The evaluation comprises three entirely independent studies that use distinct data sources, methodologies, and environments. The first study reports results from a commercial operator trial comparing LT3+\arcq\ against the native cloud-gaming stack under matched conditions. The second study analyzes an offline block-release trace captured from an Nvidia GeForce NOW session and applies the receiver-side scheduling protocol to the same underlying traffic to compare smoothed and unsmoothed release behavior. The third study evaluates \arc\ on a synthetic recovery pattern designed to reflect short-scale jitter characteristics observed in practice, which isolates the protocol behavior under controlled delay variation. Each study stands on its own, with separate data sources and experimental setups, and together they highlight different aspects of receiver-side scheduling behavior.

\subsection{Telecommunications Operator Cloud-Gaming Trial}

An evaluation was conducted in collaboration with a major telecommunications operator using a cloud-gaming workload. The operator executed controlled A/B tests comparing LT3 with \arcq\ (LT3+\arcq) against the native cloud-gaming system. The experimental setup was identical across conditions: the same cloud-gaming server, client device, access network, game title, video-encoding quality, frame rate, and scripted interaction pattern were used throughout the A/B comparison. When LT3+\arcq\ was enabled, an LT3 path was established between the client and an edge-site server, and all native traffic was carried over this path without modification to the transport protocols. When LT3+\arcq\ was disabled, traffic flowed directly between the client and the cloud-gaming server. This ensured that any performance differences were attributable solely to the presence or absence of the LT3+\arcq\ path and its integrated scheduling module.

Each set of experiments comprised multiple two-minute sessions that alternated between configurations so that both conditions experienced comparable network conditions. Tests were repeated at multiple times of day to capture naturally occurring variation in cellular and Wi-Fi channel quality. All timing and frame-delivery statistics were collected by an external analytics environment instrumenting both the client and the cloud servers.

\paragraph{Representative Video Evidence.}
Two example sessions illustrate the perceptual difference introduced by LT3+\arcq. These were back-to-back two-minute runs of a 40~Mbps, 120~fps cloud-gaming workload executed under near-identical network conditions. The first video shows the session with LT3+\arcq\ enabled \cite{brt_cloud_gaming_with}, and the second shows the native configuration \cite{brt_cloud_gaming_native}. The LT3+\arcq-enabled session exhibits visibly smoother and more consistent playback. These particular videos come from a different workload than the one summarized in Table~\ref{tab:telecom-percentiles}, but were collected under the same overall A/B methodology.

\paragraph{Representative Percentiles.}
Table~\ref{tab:telecom-percentiles} reports 95th- and 99th-percentile statistics for a representative set of eight two-minute sessions collected using a 28~Mbps, 60~fps workload. As in the larger trial design, the sessions were executed back-to-back and alternated between the two configurations, yielding four LT3+\arcq\ sessions and four Native sessions under comparable mixed cellular and Wi-Fi conditions. Each row corresponds to one session, and the rows can be interpreted as a repeated trial conducted in the same controlled experimental environment.

At 60~fps, each frame interval is approximately 16.7~ms, so percentile timing directly reflects perceptual smoothness. A deviation beyond the 95th percentile affects at least 5\% of intervals (roughly three frames per second on average), and deviations beyond the 99th percentile occur in at least 1\% of intervals. Such events often occur in bursts rather than being evenly spaced, so large excursions can create noticeable jitter clusters during gameplay.

\paragraph{Representative Percentiles.}
Table~\ref{tab:telecom-percentiles} reports 95th- and 99th-percentile statistics for a representative set of eight two-minute sessions collected using a 28~Mbps, 60~fps workload. As in the larger trial design, the sessions were executed back-to-back and alternated between the two configurations, yielding four LT3+\arcq\ sessions and four Native sessions under comparable mixed cellular and Wi-Fi conditions. Each row corresponds to one session and can be interpreted as a repeated trial conducted in the same controlled experimental environment.

At 60~fps, each frame interval is approximately 16.7~ms, so percentile timing directly reflects perceptual smoothness. A deviation beyond the 95th percentile affects at least 5\% of intervals (roughly three frames per second on average), and deviations beyond the 99th percentile occur in at least 1\% of intervals. Such events often occur in bursts rather than being evenly spaced, so large excursions can create noticeable jitter clusters during gameplay.

Table~\ref{tab:telecom-percentiles} reports these percentiles for the receiver frame interval deviation and the RTT. The receiver frame interval deviation is the difference between the observed interval between consecutive frame arrivals at the receiver and the expected frame interval of 16.7~ms. The LT3+\arcq\ configuration maintains tight receiver frame interval deviations and avoids the extreme round-trip time (RTT) outliers observed in the Native configuration. Tail latency decreases by more than a factor of three, reflecting the combined effect of LT3 and receiver-side scheduling on long delay excursions while preserving low average latency.

\begin{table}[H]
\centering
\caption{Representative percentiles for a 28~Mbps, 60~fps workload over mixed cellular and Wi-Fi paths. Comparison between the LT3+\arcq\ configuration and the Native configuration.}
\label{tab:telecom-percentiles}
\renewcommand{\arraystretch}{1.15}
\begin{tabular}{|>{\centering\arraybackslash}p{2.8cm}||
                >{\centering\arraybackslash}p{2.2cm}|
                >{\centering\arraybackslash}p{2.2cm}||
                >{\centering\arraybackslash}p{2.2cm}|
                >{\centering\arraybackslash}p{2.2cm}|}
\hline
Configuration & \multicolumn{2}{c||}{\shortstack{Receiver frame interval \\deviation [ms]}} & \multicolumn{2}{c|}{RTT [ms]} \\
\cline{2-5}
 & 95th percentile & 99th percentile & 95th percentile & 99th percentile \\
\hline \hline
LT3+\arcq & 3.4 & 3.4 & 112.0 & 132.5 \\ \hline
LT3+\arcq & 2.9 & 2.9 & 105.0 & 120.0 \\ \hline
LT3+\arcq & 2.8 & 2.8 & 107.0 & 119.0 \\ \hline
LT3+\arcq & 2.7 & 3.0 & 101.5 & 112.0 \\ \hline\hline
Native      & 13.2 & 53.6 &  47.0 & 362.0 \\ \hline
Native      & 17.2 & 69.9 &  87.0 & 424.6 \\ \hline
Native      & 12.8 & 53.8 &  61.0 & 112.5 \\ \hline
Native      & 22.6 & 76.1 & 116.6 & 469.8 \\
\hline
\end{tabular}
\end{table}

Across the trials, LT3+\arcq\ consistently reduced large jitter excursions and improved release timing without measurable throughput penalty.
Because the controller operates solely on receiver-observed recovery timing, it compensates for variable decoding and path latency on a per-block basis while leaving congestion control and coding efficiency unchanged. The results demonstrate that receiver-side scheduling provides an effective software-only mechanism for achieving tighter timing stability in field conditions.

\subsection{Offline Trace Evaluation: GeForce NOW Release Schedule}

We evaluated receiver-side scheduling on a block-release trace captured
from an NVIDIA GeForce NOW session. The trace contains the platform’s native
release timing for each block; we applied \arc/\arcq\ offline to the same
sequence to generate a smoothed release schedule.

\begin{figure}[H]
\centering
\includegraphics[width=0.9\textwidth]{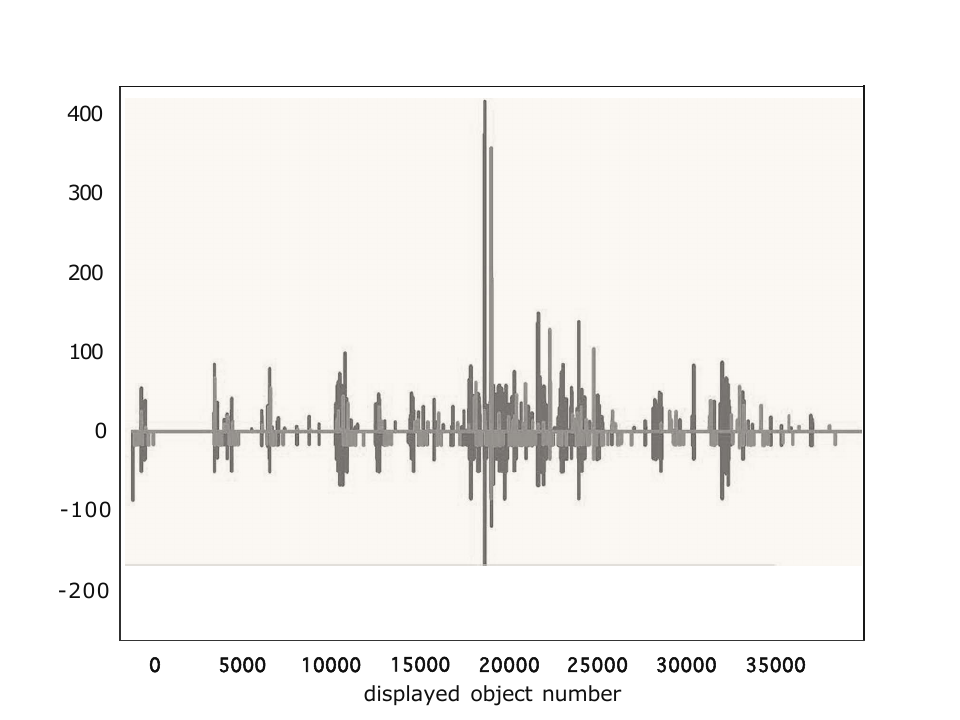}
\caption{Playback buffer deviation events for a cloud-gaming trace with and without smoothing.}
\label{fig:recbuff-jitter}
\end{figure}

Figure~\ref{fig:recbuff-jitter} shows the deviations between the release times
and a simple receiver-side playback buffer model applied to this trace.
Positive excursions represent instantaneous lags in the buffer’s output timing,
and negative excursions represent jumps.

The unsmoothed stream (dark gray trace) exhibits frequent and large excursions:
approximately 18{,}915 positive and 18{,}995 negative events over the displayed
sequence. When the same input trace is processed by the receiver-side scheduling
protocol (light gray trace), these excursions are reduced to 2{,}590 positive
and 2{,}669 negative events. The smoothed timeline therefore minimizes
oscillations in the buffer and yields a far more stable release pattern,
consistent with perceptually smooth playback under fluctuating network
conditions.

\subsection{Synthetic Recovery-Pattern Evaluation}

To isolate the effects of receiver-side scheduling under controlled timing
variability, we evaluated \arc/\arcq\ on a synthetically generated recovery
sequence. The synthetic trace was constructed to exhibit representative short-scale jitter patterns observed in practice, alternating modest delay valleys and elevated delay peaks, while preserving constant spacing between consecutive blocks at the sender.

\begin{figure}[H]
\centering
\includegraphics[width=0.85\textwidth]{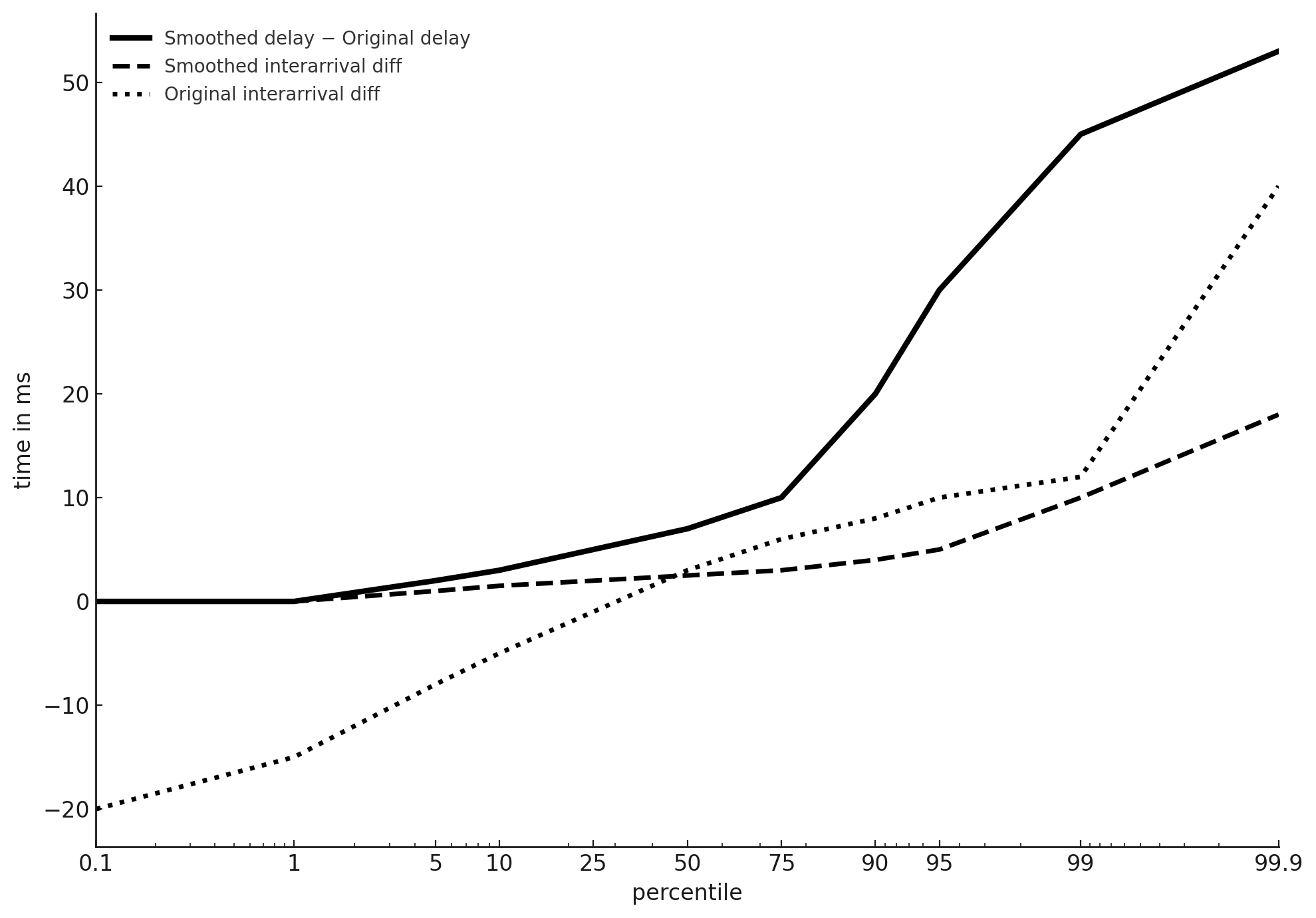}
\caption{Representative percentile comparison between smoothed and unsmoothed release timing.}
\label{fig:percentile-comparison}
\end{figure}

Figure~\ref{fig:percentile-comparison} presents a percentile comparison between
the smoothed and unsmoothed timing derived from this synthetic trace. The solid
curve shows the additional release delay introduced by the scheduler relative to
each block’s recovery time. For most blocks this added delay remains small,
increasing only at the upper percentiles where the scheduler compensates for
large positive deviations in the synthetic recovery pattern.

The dashed and dotted curves compare the inter-send time between consecutive blocks at the sender to the corresponding inter-release time at the receiver, showing how closely the smoothed schedule reproduces the original spacing between blocks at the sender.
The dotted curve corresponds to the
unsmoothed (original) recovery sequence, which exhibits wide percentile spread
and significant timing distortion. The dashed curve corresponds to the smoothed
sequence, which remains close to zero across nearly the entire
percentile range. This demonstrates that the receiver-side scheduler restores
nearly uniform spacing between releases while applying only bounded delays.

\section{Analytic Analysis and Concrete Examples}
\label{sec:analytic}

To understand the behavior of the Adaptive Release Control (\arc) scheduler, it is
valuable to analyze its evolution on simple alternating delay patterns. In this
section all quantities are normalized by setting the current peak delay to
\[
d_{\mathrm{H}} = 1,
\qquad
d_{\mathrm{L}} = 1 - g,
\qquad
0 < g < 1,
\]
and choosing the shaping scale to match the peak,
\[
U = 1.
\]
This corresponds to autoscaling the update rule so that deviations are measured
relative to the current round trip time or current one way delay. All delays
are therefore expressed in normalized units.

For each block \(n\) the LT3 sender time is \(S_n = n \cdot \tau\) for a fixed time interval \(\tau > 0\). The one way
delay alternates between the normalized peak and valley,
\[
d_n =
\begin{cases}
1,       & n \text{ even},\\[2pt]
1 - g,   & n \text{ odd},
\end{cases}
\]
so the recovery time is \(A_n = S_n + d_n\). \arc\ maintains an adaptive offset
\(D_n\) and forms the projected release time \(S_n + D_n\). The release
time is
\[
T_n = \max\{A_n,\; S_n + D_n\}.
\]
The deviation that drives the update is
\[
X_n = d_n - D_n.
\]
Throughout this section the neutral band is \(J = 0\) and the clamp \(\delta\) is large
enough to be inactive. The focus is on the fixed point behavior of the adaptive
offset when the step size \(\lambda\) is small.

\subsection{Asymptotic Analysis as \texorpdfstring{\(\lambda \to 0\)}{}}

When \(\lambda\) is small the offset moves only slightly on each iteration. In
steady state the value of \(D_n\) converges to a value \(D\) as \(\lambda \rightarrow 0\). Write the value of \(D\) as:
\[
D = 1 - \Delta,
\qquad
0 < \Delta < g,
\]
so that \(\Delta\) is the distance below the peak delay.

\paragraph{Linear asymptote (length-\(2\) pattern):} For the linear model
\[
\rho_u = \rho_\ell = 1,
\qquad
Y_n = X_n,
\]
and the update rule is
\[
D_{n+1} =
\begin{cases}
D_n + \lambda_u \cdot (1 - D_n),       & n \text{ even},\\[4pt]
D_n + \lambda_\ell \cdot ((1 - g) - D_n), & n \text{ odd}.
\end{cases}
\]

With asymmetric gains
\[
\lambda_u = (1 - \varepsilon) \cdot \lambda,
\qquad
\lambda_\ell = \varepsilon \cdot \lambda,
\qquad
0 < \varepsilon < 1,
\]
the net change across one high and one low step must be zero in steady state. Setting
\(D = 1 - \Delta\) and equating the upward and downward movements yields
\begin{equation}
    \label{eq:linear}
(1 - \varepsilon) \cdot \Delta
=
\varepsilon \cdot (g - \Delta).
\end{equation}
Solving for \(\Delta\) gives
\[
\Delta = \varepsilon \cdot g,
\qquad
D = 1 - \varepsilon \cdot g.
\]

This confirms that the linear update stabilizes at a point between the peak and
the valley, displaced \(\varepsilon \cdot g\) below the peak and
\((1 - \varepsilon) \cdot g\) above the valley. Unless \(\varepsilon\) is
extremely small, the offset remains noticeably below the upper envelope and can
never approach \(1\) closely in the \(\lambda \to 0\) regime.

\paragraph{Asymmetric exponent asymptote (length-\(2\) pattern):}
The full \arc\ update applies exponent shaping,
\[
Y_n =
\begin{cases}
X_n^{\rho_u},  & X_n > 0,\\[4pt]
- |X_n|^{\rho_\ell}, & X_n < 0,
\end{cases}
\qquad
U = 1,
\]
so on high delay steps the deviation is \(X_n = 1 - D_n\) and the change in
\(D_n\) is proportional to \((1 - D_n)^{\rho_u}\). On low delay steps the
deviation is \(X_n = (1 - g) - D_n\) and the change is proportional to
\(- (D_n - (1 - g))^{\rho_\ell}\).

Write the steady state offset as
\[
D = 1 - \Delta,
\qquad
0 < \Delta < g,
\]
so that the distance below the peak is \(\Delta\) and the distance above the
valley is \(g - \Delta\). On a high delay step the update magnitude is
\(\Delta^{\rho_u}\), and on a low delay step it is \((g - \Delta)^{\rho_\ell}\).

With asymmetric gains
\[
\lambda_u = (1 - \varepsilon) \cdot \lambda,
\qquad
\lambda_\ell = \varepsilon \cdot \lambda,
\qquad
0 < \varepsilon < 1,
\]
the net change in \(D\) across one high and one low step in the
\(\lambda \to 0\) regime must be zero at the fixed point. 
The balance condition is
\[
(1 - \varepsilon) \cdot \lambda \cdot \Delta^{\rho_u}
=
\varepsilon \cdot \lambda \cdot (g - \Delta)^{\rho_\ell},
\]
which is equivalent to Equation~(\ref{eq:linear}) in the special case
when \(\rho_u = \rho_\ell = 1\).
Canceling \(\lambda > 0\) and rearranging gives the fixed point equation
\[
\Delta^{\rho_u}
=
\frac{\varepsilon}{1 - \varepsilon} \cdot (g - \Delta)^{\rho_\ell}.
\]

This equation has a unique solution in \((0, g)\). It also yields a useful
upper bound on \(\Delta\). Since \(0 < g - \Delta < g\), we have
\[
(g - \Delta)^{\rho_\ell} < g^{\rho_\ell},
\]
and therefore
\[
\Delta^{\rho_u}
=
\frac{\varepsilon}{1 - \varepsilon} \cdot (g - \Delta)^{\rho_\ell}
<
\frac{\varepsilon}{1 - \varepsilon} \cdot g^{\rho_\ell}.
\]
Taking the \(\rho_u\)th root gives
\[
0 < \Delta
<
\left(\frac{\varepsilon}{1 - \varepsilon}\right)^{1 / \rho_u}
\cdot g^{\rho_\ell / \rho_u}.
\]

Thus the steady state offset satisfies
\[
D = 1 - \Delta
>
1
-
\left(\frac{\varepsilon}{1 - \varepsilon}\right)^{1 / \rho_u}
\cdot g^{\rho_\ell / \rho_u}.
\]

For the exponent asymmetry of interest, with \(0 < \rho_u < 1 < \rho_\ell\) and \(g < 1\), the exponent \(\rho_\ell / \rho_u\) is greater than one, which makes \(g^{\rho_\ell / \rho_u}\) very small. In typical operating regimes \(\varepsilon < 1/2\), so \(\varepsilon/(1 - \varepsilon)\) is less than one. Since \(1/\rho_u > 1\), raising this ratio to the power \(1/\rho_u\) reduces it even further. Consequently,
\[
\left(\frac{\varepsilon}{1 - \varepsilon}\right)^{1 / \rho_u}
\cdot g^{\rho_\ell / \rho_u}
\]
is extremely small, and the equilibrium offset \(\Delta\) lies very close to the peak value \(d_{\mathrm H} = 1\).

In the special case \(\varepsilon = \tfrac{1}{2}\) this bound reduces to
\[
\Delta < g^{\rho_\ell / \rho_u},
\]
which matches the symmetric step size result.

\subsection{Why \texorpdfstring{\(\lambda\)}{} and \texorpdfstring{\(\varepsilon\)}{} Cannot Be Too Small}

Although the \(\lambda \to 0\) regime yields clean analytical expressions and sharp upper envelope tracking under exponent asymmetry, choosing \(\lambda\) too small is undesirable in practice. The rate at which the offset approaches its steady state is governed by a contraction whose speed is proportional to \(\lambda\), so the convergence time grows on the order of \(1/\lambda\). In realistic network settings the path delay can fluctuate or the sending rate can shift on short timescales, and very small values of \(\lambda\) cause the scheduler to react too slowly to such global changes.

Similarly, \(\varepsilon\) cannot be chosen too small. When blocks are consistently recovered early after an upward fluctuation in the projected release time, the downward correction is driven by the low step with gain \(\lambda_\ell = \varepsilon \cdot \lambda\). If \(\varepsilon\) is too small, the projected release time decreases only very slowly, causing extended periods during which early recoveries do not meaningfully reduce the offset. This slows the reaction to downward trends in recovery times and reduces the scheduler’s ability to realign with the upper envelope.

In general there is a tradeoff between \(\lambda\) and \(\varepsilon\). Smaller \(\lambda\) slows global responsiveness, while smaller \(\varepsilon\) slows downward adjustments following early recoveries. Both parameters should therefore be kept large enough to ensure timely reactions to changing conditions, while still small enough to smooth the inter-release times of blocks and maintain stable release timing.

\subsection{Concrete Examples}

To illustrate the behavior in realistic units, we interpret
\(d_{\mathrm H} = 1.0\) as a delay of \(100\,\mathrm{ms}\) and
\(d_{\mathrm L} = 0.7\) as \(70\,\mathrm{ms}\), so the gap is
\[
g = d_{\mathrm H} - d_{\mathrm L} = 30\,\mathrm{ms}.
\]
We also take the inter-send interval to be
\[
\tau = 16.7\,\mathrm{ms},
\]
corresponding to a \(60\,\mathrm{fps}\) video stream in which each block is a
frame. In all concrete examples below we fix
\[
\lambda = 0.2,
\qquad
\varepsilon = 0.2,
\]
initialize the offset according to the first recovered block,
\[
D_0 = d_{\mathrm H} = 1.0 \longrightarrow 100\,\mathrm{ms},
\]
and then iterate the update rules for a long sequence of blocks until the
behavior becomes periodic. For the exponent-asymmetry cases we use the same
pair of exponents throughout:
\[
\rho_u = 0.5,
\qquad
\rho_\ell = 2.0.
\]

\paragraph{Linear example (length-\(2\) pattern):}
For the alternating pattern \(S = 2\) the nominal delays are
\[
d_n =
\begin{cases}
100\,\mathrm{ms}, & n \text{ even},\\[2pt]
70\,\mathrm{ms},  & n \text{ odd}.
\end{cases}
\]
The gains are
\[
\lambda_u = (1 - \varepsilon) \cdot \lambda = 0.16,
\qquad
\lambda_\ell = \varepsilon \cdot \lambda = 0.04.
\]

Running the linear update with the sign-aware rule
\[
D_{n+1}
=
\begin{cases}
D_n + \lambda_u \cdot (100\,\mathrm{ms} - D_n), & X_n > 0,\\[4pt]
D_n + \lambda_\ell \cdot (70\,\mathrm{ms} - D_n),  & X_n < 0,
\end{cases}
\qquad
X_n = d_n - D_n,
\]
produces a stable two-point cycle for the effective offset at high and low
steps. In steady state the values are approximately
\[
D_{\mathrm H} \approx 93.8\,\mathrm{ms},
\qquad
D_{\mathrm L} \approx 94.8\,\mathrm{ms}.
\]
On high-delay steps the release follows the arrival,
\[
T_n - S_n = 100\,\mathrm{ms},
\]
while on low-delay steps the release is governed by the offset,
\[
T_n - S_n \approx 94.8\,\mathrm{ms}.
\]

The inter-release intervals therefore alternate between
\[
T_n - T_{n-1} \approx 21.9\,\mathrm{ms}
\quad\text{and}\quad
T_n - T_{n-1} \approx 11.5\,\mathrm{ms}.
\]
The linear scheduler produces strongly uneven release timing at the receiver, with
every other interval substantially shorter than \(\tau\) and the others
substantially longer.

\paragraph{Asymmetric exponent example (length-\(2\) pattern):}
With exponent shaping
\[
\rho_u = 0.5,
\qquad
\rho_\ell = 2.0,
\]
and the same \(\lambda\) and \(\varepsilon\), the update becomes
\[
D_{n+1}
=
\begin{cases}
D_n + \lambda_u \cdot (X_n)^{\rho_u}, & X_n > 0,\\[4pt]
D_n - \lambda_\ell \cdot (-X_n)^{\rho_\ell}, & X_n < 0,
\end{cases}
\qquad
X_n = d_n - D_n.
\]

Starting from \(D_0 = 100\,\mathrm{ms}\) and iterating, the system converges to a
short-period cycle in which the effective delays
\[
T_n - S_n = \max\{d_n,\; D_n\}
\]
remain very close to the peak. In steady state the releases for both high and
low steps lie in a narrow band between approximately
\[
97.9\,\mathrm{ms}
\quad\text{and}\quad
100.5\,\mathrm{ms},
\]
with low-delay arrivals occasionally being held slightly past the nominal peak
due to overshoot in the offset.

The resulting inter-release intervals follow a four-element pattern in steady
state:
\[
T_n - T_{n-1}
\approx
\{16.3,\, 16.7,\, 16.5,\, 17.2\}\,\mathrm{ms},
\]
repeating. All intervals stay close to the nominal \(\tau = 16.7\,\mathrm{ms}\),
and the jitter is limited to about \(\pm 0.5\,\mathrm{ms}\). The exponent
asymmetry therefore yields almost uniform release timing, in contrast to
the pronounced two-point oscillation under the linear update.

\subsection{Extension to longer recovery patterns}

The alternating pattern analyzed above corresponds to pattern length 
\(2\), in 
which each late recovery is immediately followed by an early recovery. In 
realistic environments, however, late recoveries often appear singly, while 
early recoveries occur in consecutive sequences. This motivates extending the 
analysis to a length-\(P\) pattern in which one block has peak delay 
\(d_{\mathrm H} = 1\) and the remaining \(P - 1\) blocks have valley delay 
\(d_{\mathrm L} = 1 - g\). The pattern then repeats every \(P\) steps.

As before, write the steady-state offset in the small-\(\lambda\) regime as
\[
D = 1 - \Delta,
\qquad
0 < \Delta < g.
\]

\paragraph{Linear asymptote (length-\(P\) pattern):}  
With asymmetric gains
\[
\lambda_u = (1 - \varepsilon) \cdot \lambda,
\qquad
\lambda_\ell = \varepsilon \cdot \lambda,
\]
and symmetric exponents \(\rho_u = \rho_\ell = 1\), the fixed point satisfies
\[
(1 - \varepsilon) \cdot \Delta
=
(P - 1) \cdot \varepsilon \cdot (g - \Delta).
\]
Solving for \(\Delta\) gives the linear steady-state distance below the peak:
\[
\Delta
=
\frac{(P - 1) \cdot \varepsilon}{1 + (P - 2) \cdot \varepsilon} \cdot g,
\qquad
D = 1 - \Delta.
\]
This reduces to \(\Delta = \varepsilon \cdot g\) when \(P = 2\). As 
\(P\) increases, the influence of the valley steps accumulates and the fixed 
point moves further into the valley.

\paragraph{Asymmetric exponent asymptote (length-\(P\) pattern):}  
With exponent shaping,
\[
Y_n =
\begin{cases}
X_n^{\rho_u}, & X_n > 0,\\[3pt]
- |X_n|^{\rho_\ell}, & X_n < 0,
\end{cases}
\qquad
0 < \rho_u < 1 < \rho_\ell,
\]
the balance over one period becomes
\[
(1 - \varepsilon) \cdot \Delta^{\rho_u}
=
(P - 1) \cdot \varepsilon \cdot (g - \Delta)^{\rho_\ell}.
\]
This equation has a unique solution in \((0, g)\). A convenient upper bound 
follows from \(0 < g - \Delta < g\):
\[
\Delta
<
\left(
\frac{(P - 1) \cdot \varepsilon}{1 - \varepsilon}
\right)^{1 / \rho_u}
\cdot g^{\rho_\ell / \rho_u},
\qquad
D = 1 - \Delta.
\]
This bound is informative only when the multiplicative factor 
\((P - 1) \cdot \varepsilon / (1 - \varepsilon)\) is less than one.  
Solving 
\[
\frac{(P - 1) \cdot \varepsilon}{1 - \varepsilon} < 1
\]
gives the condition \(\varepsilon < 1/P\).  
When \(\varepsilon\) satisfies this inequality, the right-hand side of the bound 
is strictly smaller than \(g\), and the quantity
\[
\left( \frac{(P - 1) \cdot \varepsilon}{1 - \varepsilon} \right)^{1/\rho_u} 
\cdot g^{\rho_\ell/\rho_u}
\]
becomes very small in the regime \(0 < \rho_u < 1 < \rho_\ell\), implying that 
\(\Delta\) lies close to the peak. When \(\varepsilon \ge 1/P\), the bound no 
longer provides a meaningful guarantee.

These expressions generalize the \(P = 2\) results and show that exponent 
asymmetry continues to keep the offset near the peak even when many consecutive 
valley recoveries occur, provided \(\varepsilon\) is not too large and 
\(g^{\rho_\ell / \rho_u}\) remains small.

\subsection{Concrete example for length-10 pattern}

To illustrate the effect of an extended recovery pattern, consider the case
\(P = 10\), corresponding to one late recovery followed by nine early
recoveries. We again interpret
\[
d_{\mathrm H} = 1.0 \longrightarrow 100\,\mathrm{ms},
\qquad
d_{\mathrm L} = 0.7 \longrightarrow 70\,\mathrm{ms},
\qquad
g = 30\,\mathrm{ms},
\]
and use the same inter-send interval
\[
\tau = 16.7\,\mathrm{ms},
\]
the same gains \(\lambda = 0.2\), \(\varepsilon = 0.2\), and the same
initialization \(D_0 = 100\,\mathrm{ms}\). One block in each 10-block period
has delay \(100\,\mathrm{ms}\); the remaining nine have delay \(70\,\mathrm{ms}\).

\paragraph{Linear example (length-\(10\) pattern):}
As before, the gains on upward and downward corrections are
\[
\lambda_u = (1 - \varepsilon) \cdot \lambda = 0.16,
\qquad
\lambda_\ell = \varepsilon \cdot \lambda = 0.04.
\]
On each step the update
\[
D_{n+1}
=
\begin{cases}
D_n + \lambda_u \cdot (100\,\mathrm{ms} - D_n), & X_n > 0,\\[4pt]
D_n + \lambda_\ell \cdot (70\,\mathrm{ms} - D_n),  & X_n < 0,
\end{cases}
\qquad
X_n = d_n - D_n,
\]
is applied using the sign of the deviation \(X_n\), regardless of whether the
nominal step is a peak or a valley.

After a long warm-up, the effective delays \(T_n - S_n = \max\{d_n, D_n\}\) 
stabilize into a 10-point cycle. In one representative period the values are
\[
T_n - S_n
\approx
\{100.0,\, 81.5,\, 81.0,\, 80.6,\, 80.3,\,
79.9,\, 79.6,\, 79.2,\, 78.8,\, 78.3\}\,\mathrm{ms},
\]
where the first element corresponds to the late step and the remaining nine to
the early steps. The offset continues to drift downward over the early steps,
so valley blocks are released progressively closer to the raw \(70\,\mathrm{ms}\)
arrival.

The inter-release intervals over the same 10-block period are
\[
T_n - T_{n-1}
\approx
\{38.4,\; -1.8,\; 16.2,\; 16.3,\; 16.3,\;
16.3,\; 16.3,\; 16.3,\; 16.3,\; 16.4\}\,\mathrm{ms}.
\]
The negative interval indicates that one block is scheduled to be released
slightly \emph{before} its predecessor, so the raw update rule can generate
out-of-order releases. In steady state the late block induces a long
inter-release gap of almost \(40\,\mathrm{ms}\), more than
\(20\,\mathrm{ms}\) above the nominal inter-send interval
\(\tau = 16.7\,\mathrm{ms}\); this large lag then forces the next block to be
scheduled roughly \(2\,\mathrm{ms}\) before its predecessor, with the remaining
eight intervals clustered near \(16.3\,\mathrm{ms}\).

\paragraph{Exponent asymmetry example (length-\(10\) pattern):}
With exponent shaping
\[
\rho_u = 0.5,
\qquad
\rho_\ell = 2.0,
\]
and the same \(\lambda\) and \(\varepsilon\), the update is
\[
D_{n+1}
=
\begin{cases}
D_n + \lambda_u \cdot (X_n)^{\rho_u}, & X_n > 0,\\[4pt]
D_n - \lambda_\ell \cdot (-X_n)^{\rho_\ell}, & X_n < 0,
\end{cases}
\qquad
X_n = d_n - D_n.
\]

In steady state the effective delays again form a 10-point cycle. A
representative period has
\[
T_n - S_n
\approx
\{100.0,\, 99.6,\, 99.3,\, 98.9,\, 98.6,\,
98.3,\, 98.0,\, 97.6,\, 97.3,\, 97.0\}\,\mathrm{ms},
\]
with the first value associated with the late step and the others with the nine
early steps. Despite each early recovery arriving \(30\,\mathrm{ms}\) before the
peak, the projected release times for valley blocks stay within a few
milliseconds of the peak.

The corresponding inter-release intervals over the same period are
\[
T_n - T_{n-1}
\approx
\{19.7,\, 16.3,\, 16.3,\, 16.4,\, 16.4,\,
16.4,\, 16.4,\, 16.4,\, 16.4,\, 16.4\}\,\mathrm{ms}.
\]
Aside from a single interval at the start of each period that is extended to
about \(19.7\,\mathrm{ms}\) (a modest additional delay of roughly
\(3\,\mathrm{ms}\) beyond \(\tau\)), all intervals lie very close to
\(\tau = 16.7\,\mathrm{ms}\), with variations of at most a few tenths of a
millisecond. There are no negative intervals, so the release order is preserved
without any additional guard.

Even with nine consecutive early recoveries per cycle, exponent asymmetry 
maintains tight upper-envelope tracking and a nearly uniform release timing.
In contrast, the linear update rule produces large oscillations and can 
schedule blocks out of order, with the severity of the pattern increasing 
as \(P\) grows.

\section{Discussion and Limitations}
The asymmetric response can retain residual latency after a delay spike, since it decreases more slowly during the subsequent valley; the clamp parameter \(\delta\) limits how much of this residual drift can accumulate.
The quantization step \(\gamma\) should be commensurate with the spacing of data arrivals at the sender to avoid visible stepping. The guard \(G\) improves safety for in-order transports at the cost of small delays when predecessors are late.

Receiver-side scheduling is complementary to transport-level mechanisms. It does not replace congestion control, retransmission, or existing jitter buffers, but instead adds a timing layer that operates on recovered blocks. Exploring joint design with transport protocols and automated parameter selection based on observed trace statistics are natural extensions of this work.

\section{Conclusion}
Receiver-side scheduling stabilizes block release timing with minimal complexity and without sender coordination. The adaptive offset, quantized hysteresis, and bounded in-order guard together provide low-jitter, low-latency behavior across variable network conditions. Experiments with a cloud-gaming workload show that integrating \arc\ into LT3 resolves virtually all large jitter excursions and significantly tightens inter-release timing. These results suggest broader applicability to interactive streaming and related workloads that require tight timing at the receiver.

\section*{Acknowledgments}

This material is based upon work supported in part by the National Science Foundation under Award~2212574.

\section*{Intellectual Property Notice}

This document describes technology that may be covered by issued or pending patents owned by BitRipple, Inc. 
No license to such patents is granted by this publication.

\bibliographystyle{plain}
\bibliography{references}

\end{document}